\documentclass[prb,aps,amssym,nofootinbib,floatfix,
twocolumn,notitlepage]{revtex4-1} 

\usepackage{graphicx}
\usepackage{xcolor}
\usepackage{amsmath,amssymb,bbold,bm,amsfonts}
\usepackage[font=small,justification=justified]{caption}
\usepackage{cancel}
\usepackage{comment}

\newcommand{\be}{\begin{equation}}
\newcommand{\een}{\end{equation*}}
\newcommand{\bs}{\begin{split}}
\newcommand{\ben}{\begin{equation*}}
\newcommand{\ee}{\end{equation}}
\newcommand{\es}{\end{split}}
\newcommand{\bmx}{\begin{array}}
\newcommand{\emx}{\end{array}}
\newcommand{\bea}{\begin{eqnarray}}
\newcommand{\bean}{\begin{eqnarray*}}
\newcommand{\eea}{\end{eqnarray}}
\newcommand{\eean}{\end{eqnarray*}}
\newcommand{\dg}{^{\dagger}}
\newcommand{\dn}{^{\vphantom{\dagger}}}

\newcommand{\da}{\downarrow}
\newcommand{\bb}[1]{\mathbb{#1}}

\newcommand{\eps}{\epsilon}

\newcommand{\sgn}[1]{{\rm sign}{#1}}
\newcommand{\pref}[1]{(\ref{#1})}

\newcommand{\intinf}[1]{\int_{-\infty}^{+\infty}{#1}}

\newcommand{\intob}[1]{\int_{0}^{\beta}{#1}}

\newcommand{\im}[1]{{\rm Im}\left[ #1 \right]}
\newcommand{\tr}[1]{{\rm Tr}\Big[ #1 \Big]}

\newcommand{\abs}[1]{\left\vert #1 \right\vert}
\newcommand{\bra}[1]{\left\langle #1 \right\vert}
\newcommand{\ket}[1]{\left\vert #1\right\rangle}
\newcommand{\braket}[1]{\left\langle #1\right\rangle}

\newcommand{\com}[2]{\left[#1,#2\right]}
\newcommand{\acom}[2]{\left\{#1,#2\right\}}
\newcommand{\mat}[1]{\left(\bmx{cc}#1\emx\right)}

\newcommand{\bw}[1]{\begin{widetext}}
\newcommand{\ew}[1]{\end{widetext}}

\newcommand{\red}[1]{{\color{red} #1}}
\newcommand{\gray}[1]{}

\begin{document}
\title{Majorana approach to the stochastic theory of lineshapes}
\author{Yashar Komijani$^{1, *}$, and Piers Coleman$^{1}$}
 \affiliation{ $^1$Department of Physics and Astronomy, Rutgers University, Piscataway, New Jersey, 08854, USA}
\date{\today}
\begin{abstract}
Motivated by recent M\"ossbauer experiments on strongly correlated
mixed-valence systems, we revisit the Kubo-Anderson stochastic
theory of spectral lineshapes. 
Using a Majorana representation for the
nuclear spin we demonstrate how to recast the 
classic lineshape theory in a 
field-theoretic and diagrammatic language. 
We show that
the leading contribution to the self-energy can reproduce most of the
observed lineshape features including splitting and lineshape
narrowing, while the vertex and the self-consistency corrections can
be systematically included in the calculation. This new approach
permits us to predict the line-shape produced by an 
arbitrary bulk charge fluctuation spectrum providing a
model-independent way to extract the local charge fluctuation spectrum
of the surrounding medium. We also derive an inverse formula to extract the charge fluctuation from the measured lineshape.
\end{abstract}
\maketitle

\section{Introduction}

Resonant spectroscopy of two-level systems is a powerful tool to study
the environment in which they are embedded, and include photo and
$X$-ray absorption, electron spin resonance, nuclear magnetic
resonance and even current spectroscopy of a quantum dot. An important
example is M\"ossbauer spectroscopy,\,\cite{Mossbauer} where nuclear
transitions in a solid are studied by the recoil-free resonant
absorption/emission of $\gamma$-ray photons, providing a 
sensitive probe of low frequency 
electric charge fluctuations and the magnetic field
texture in the material. Recently, the advent of
synchrotron-based-radiation as a new source of hard
$X$-rays\,\cite{Seto09} has opened up a wide range of materials 
to  M\"ossbauer study, with the possibility of new insights
into strongly correlated systems, such as
YbAlB$_4$.\,\cite{Kobayashi}

In spite of this general applicability and recent demand, the stochastic theory
of lineshape has been largely unchanged since the seminal works of
Anderson, Kubo and later Blume on the topic.\,\cite{Anderson54,Kubo54,Kubo54b,Blume68,Blume68b,Cianchi86} The
classic theory is model-based, 
providing information about the lineshape in
environments with specialized Gaussian or Markovian dynamics.
This motivates us to revisit the problem of spectral lineshapes in two level
systems, recasting the problem in a modern framework. 
Our work leads us to conclude that there 
is a wide parameter regime in M\"ossbauer spectroscopy, in which the entanglement
between two-level system and the environment is weak enough for the measured absorption lineshape to provide model-independent
information about the spectrum of charge fluctuations in the 
surroundings.  

\begin{figure}[ht!]
\includegraphics[width=1\linewidth]{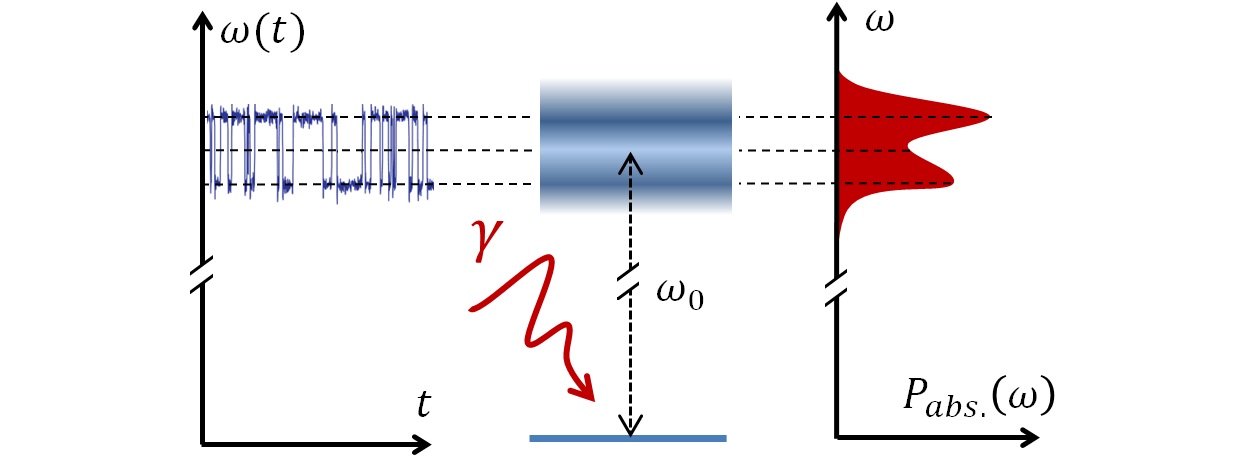}
\caption{\raggedright\small The setup considered in this paper. The resonance frequency of the two-level system as probed by gamma ray absorption (middle panel) is modulated by the quantum/thermal fluctuations of an additional degree of freedom (left panel), resulting in a modification of the spectral lineshape (right panel). The average resonance frequency $\omega_0$ is much larger than any other energy scale in the problem and the temperature is low enough so that the two-level system is mostly at its ground state.
}\label{fig0}
\end{figure}

\section{The model}

Here, we study the spectroscopic lineshape of a two-level system
(probe) immersed in a fluctuating 
environment [Fig.\,\pref{fig0}]. 
Quite generally, the Hamiltonian of a two-level system can be
describes by a pseudo-spin variable $I_z=\pm 1/2$ in a 
Zeeman field which is modulated by another degree of freedom
\be H=(\omega_0+\lambda \sigma_f)(I^z+1/2)+H_{\rm
rest}\{\sigma_f\}.\label{eq1} \ee The level spliting is modified 
by the fluctuating bosonic variable $\sigma_f$ whose dynamics
is governed by $H_{\rm rest}$. $I^{z}$ commutes with both $H_{\rm
rest}$ and $\sigma_{f}$ but the latter two do not commute, resulting in resonance-frequency fluctuations. 
A main assumption is that the level-splitting
{(which
corresponds to a Nuclear excitation energy in M\" ossbauer spectroscopy)} is 
large enough ($\omega_0\gg T$) so that the two-level system is in its ground state.
Without much restriction, we assume that $\braket{\sigma_f}=0$ and its average is absorbed in $\omega_0$. Most of our discussion
is applicable to arbitrary $\sigma_f$ and $H_{\rm rest}$ and the latter can be quite complicated. 
However, we will make a
distinction between the cases when $\sigma_f$ is a continuous variable
and when it is has discrete eigenvalues. The latter occurs, for
example when $\sigma_f=\pm 1$ is a digital variable caused by the
capacitive coupling of the probe to the occupation of a nearby
fermionic $f$-level. On the other hand, if the probe is coupled to
many independent $f$-levels, the average becomes a continuous variable
which according to the central limit theorem, will have Gaussian dynamics.

In spectroscopic studies, the above system is coupled to an additional
photonic degree of freedom, $H'=H+g(I^+a+h.c.)+\omega a\dg a$ which is
used to probe the line-shape of the excited state at absorption
energies $\omega\sim \omega_{0}$. The absorbed power (positive) is
written (see app.\,1) in terms of the retarded Green's function of the
spins \be
P(\omega)=-\omega{\Omega^2}\chi''_{-+}(\omega+i\eta),\label{eq2}
\ee where $\chi_{-+}''(\omega+i\eta)$ denotes the imaginary part of
the Fourier transform of the retarded function
$\chi_{-+}^R(t)=-i\theta(t)\braket{\com{I^-(t)}{I^+(0)}}$, evaluated
in thermal equilibrium, and $\Omega^{2}$ is the field strength of the
incoming radiation. 


\section{Majorana representation and the self-energy diagrams}

To use the machinery of field theory, we need to represent the
spin in terms of canonical fields.  We use a
Majorana representation of the spin,\,\cite{Mao03,Schnirman} $\vec
I=-\frac{i}{2}\vec\eta\times\vec\eta$ which expresses the spin in
terms of three\,\cite{footnote2} Majorana fermions $\eta^i$, $i=1,2,3$
obeying the anti-commutation relation
$\acom{\eta^i}{\eta^j}=\delta^{ij}$. The advantage of this
representation, is that it avoids the use of a constraint
and furthermore, 
the spin dynamics can be directly read-off from the {\it one-particle} 
Green's function of the Majorana fermions.\,\cite{Mao03,Schnirman} This is
because the Majorana composite $\phi = -2 i \eta_{1}\eta_{2}\eta_{3}$
commutes with the spin $[\phi ,\vec{I}]=0$ and the Hamiltonian
$[\phi , H]=0$, and 
is  thus a constant of motion. Moreoever, since 
$2 \phi \vec{I}= \vec{\eta }$ and $\phi^{2}=1/2$, it
follows that 
\be\label{keyequation}
\langle I^{a}
(t)I^{b} (t')\rangle =\frac{1}{2}\langle  \eta^{a} (t)\eta^{b} (t')\rangle  .
\ee
It is convenient
to combine  two of the Majoranas into a single Dirac fermion,
$
d\dg\equiv (\eta^1+i\eta^2)/\sqrt 2 $, so that $I^z= d\dg d-1/2$ and
$I^+=\sqrt 2\eta^3d\dg$. 
The occupied/unoccupied states of the
$d$-level correspond to the up/down states of the probe isospin, $I_{z}$. 
By Eq.\,(\ref{keyequation})
$I^-(t)I^+=d(t)d\dg$ and $I^+I^-(t)=d\dg d(t)$ are true at all times
and therefore,
$\braket{I^-(t)I^+(0)-I^+(0)I^-(t)}=\braket{d(t)d\dg(0)-d\dg(0)d(t)}$. Fourier
transform of the left side is the imaginary part of the spin
susceptibility, whereas the right-side has the `wrong' sign to be a
fermionic retarded function. Instead, it is the Keldysh function of
the $d$-level, and
by the fluctuation-dissipation theorem can be related to the imaginary
part of the retarded
function. 
Therefore, we
obtain\,\cite{Mao03} (see app.\,2) 
\begin{equation}\label{eq3}
\chi''_{-+}(\omega+i\eta)=\tanh(\beta\omega/2){G''_d(\omega+i\eta)}.
\end{equation}
Here, $\beta=1/T$ is the inverse temperature and
$G''_d(\omega+i\eta)$ is the imaginary part of the Fourier transform
of the retarded Greens function, $G^R_d(t)=
\braket{-i\theta(t)\acom{d(t)}{d\dg(0)}}$. Quite generally, we expect
the absorption function, 
$\propto G''_d(\omega+i\eta)$ to be 
narrow function centered at $\omega_0$, so combining Eqs\,\pref{eq2} and
\pref{eq3}, we have $P(\omega)\propto -G''_d(\omega+i\eta)$ and the
proportionality constant is
$c\approx\omega_0\tanh(\beta\omega_0/2)\Omega^2$. The area under the resonance is constant, leading to the sum rule $\int{{d\omega}
P(\omega)}=-c\int{{d\omega}}G''_d(\omega+i\eta)=\pi
c\acom{d}{d\dg}=\pi c$.

\begin{figure}[t] 
\includegraphics[width=1\linewidth]{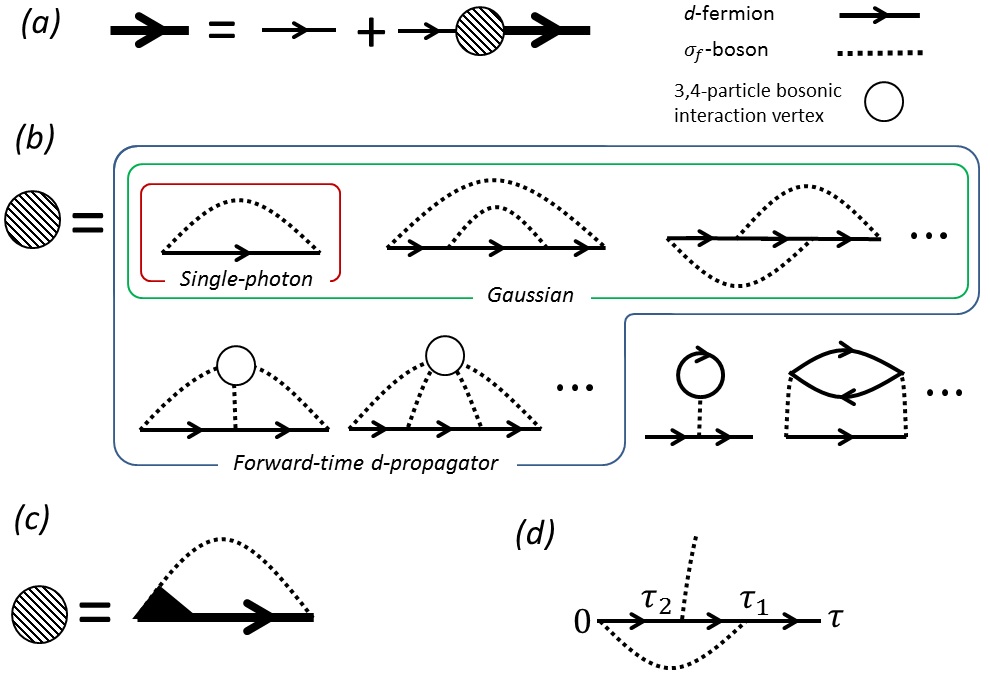}
\caption{\raggedright\small (color online) (a) Diagrammatic representation of the Dyson's equation. (b) Perturbative expansion of the self-energy to order $O(\lambda^4)$. The set of diagrams corresponding to the (leading) single-photon approximation, the Gaussian approximation and the forward-time $d$-propagator approximations are indicated. (c) The self-energy can be obtained self-consitently from the leading-order by including the vertex correction and upgrading the $d$-propagator to a full Green's function. (d) The leading contribution to the vertex correction, used in the text.}\label{fig:Fig1}
\end{figure}

\gray{So far, everything was exact.}  
The advantage of the Majorana
representation is that one can apply standard field theory
techniques, developing a Feynman expansion for the one-particle 
Green's function $G_d(\tau)\equiv-\braket{T_\tau
d(\tau)d\dg(0)}$, which we can immediately convert to a spin
correlation function of the probe isospin using Eq.\,\pref{eq3}. Taking advantage of the 
Dyson's equation for the Green's
function, $G_{d} (z) =
[z-\omega_{0}-\Sigma_d(z)]^{-1}$, the $d$-fermion Green's function
is described in terms of the self-energy $\Sigma_d(z)$ of the $d$-fermion (Fig.\,\ref{fig:Fig1}a). 
The calculation of the spin dynamics then reverts to a calculation
of the self-energy of the $d$ fermion. The relevant diagrams in the expansion of the self-energy to order $O(\lambda^4)$ are shown in Fig.\,\ref{fig:Fig1}b. Assuming
that temperature is much smaller than $\omega_0$, we have
$f(\omega_0)\approx 0$, where $f(\omega)=[e^{\beta\omega}+1]^{-1}$ is
the Fermi-Dirac distribution. Therefore, to a good approximation, it
suffices to consider an exclusively 
forward-time bare $d$-level
propagator
$g_d(\tau)=[-\theta(\tau)+f(\omega_0)]e^{-\omega_0\tau}\approx
-\theta(\tau)e^{-\omega_0\tau}$ in the calculations. This
considerably simplifies the diagrams, because it
eliminates all diagrams that involve fermions propagating backwards in
time, i.e any diagrams with additional fermion
loops (e.g. the last two Feynman diagrams of Fig.\,(2b)).
Physically, this approximation means that there is no back-action from the $d$-level on the charge fluctuations, e.g. by the last diagram in Fig.\,(2b). The forward-time restriction on the $d$-propagator limits us 
to a single $d$-fermion branch in the Green's function, 
and these diagrams re-sum to\,\cite{Anderson54}
\be
G_d(\tau)=-e^{-\omega_0\tau}\braket{T_\tau e^{\lambda\int_0^\tau d\tau'\sigma_f(\tau')}}. \label{eqExact}
\ee
Alternatively, this formula can be obtained using the methods similar to the orthogonality catastrophe problem by writing $d(\tau)=e^{\tau H_-}de^{-\tau H_+}$ where $H_{\pm}$ correspond to the $I_z=\pm 1/2$ sectors of the Hamiltonian, respectively. 

Expanding the exponent inside the bracket of Eq.\,\pref{eq3}, we have
to evaluate $n$-point correlation functions of $\sigma_f$. These have
a disconnected part (sum of all possible Wick's contractions, the Gaussian subset in Fig.\,(2b)) and a
connected part caused by the interaction vertices of $H_{\rm rest}$. For a continuous $\sigma_f$ variable with Gaussian
dynamics, the connected-part contributions to Eq\,\pref{eqExact} is
zero and the bracket becomes
$\exp[{\frac{1}{2}\lambda^2\int_0^\tau{d\tau_1}\int_0^\tau{d\tau_2}\braket{T_\tau
\sigma_f(\tau_1)\sigma_f(\tau_2)}}]$ (see app.\,3).  \gray{which can
also be obtained from resuming the Wick-contracted diagrams} We can
simplify this to write\,\cite{Anderson54}
\bea
G_d(\tau)\to -e^{-\omega_0\tau}\exp\Big[{-\lambda^2\int_0^{\tau}dx(\tau-x)\chi_C(x)}\Big].\label{eqGaussian}
\eea
Here, $\chi_C(\tau)=-\braket{T_\tau \sigma_f(\tau)\sigma_f(0)}$ is the
correlation function of the fluctuations and we used that
$\chi_C(-\tau)=\chi_C(\tau)$.  This formula, due to
Anderson\,\cite{Anderson54} 
is quite
precise within the Gaussian-action assumption, but unfortunately due
to the appearance of the $\chi_C$ within an integral in the exponent,
it is difficult to use it to extract the susceptibility
$\chi_C(\omega)$ from the spectrum.

On the other hand, for a discrete $\sigma_f=\pm 1$ variable we
cannot use Anderson's Gaussian formula due to the importance of interaction vertices. For example, replacing the composite charge bosons by the population of a fermionic $f$-level $\sigma_f=2f\dg f-1$, we see that $H_{\rm rest}$ can be strongly interacting with $O(1)$ vertex corrections.  In this case, the resumation of the Gaussian
subset of diagrams to infinite order is too arbitrary.  A special case
in which the discrete problem can be solved, is when the discrete $\sigma_f$ has
(classical) Markovian dynamics, i.e. its probability of being $\pm 1$ at any given
time, $\vec p(t)=\mat{p_+ & p_-}^T$, obeys a rate equation $\dot{\vec
p}=\bb\Gamma \vec p$ with the transition matrix $\bb\Gamma$. Interpreting Eq.\,\pref{eqExact} as a statistical average, we follow Anderson\,\cite{Anderson54} and divide
the integral in Eq.\,\pref{eqExact} to $N\to\infty$ segments and use the rate
equation to derive (see app.\,4) \be G_d(\tau)\to-e^{-\omega_0\tau}
\{\vec 1\hphantom{ .}^T.\exp[\tau(\lambda\sigma^z-i\bb\Gamma)].\vec
p_{st}\}.\label{eqDM} \ee Here, $\sigma^z$ is in the same space as
$\bb\Gamma$. The vector $\vec p_{st}$ contains steady-state
probabilities and is the solution to $\bb\Gamma\vec p_{st}=0$. Much of
the stochastic theory is about diagonalizing the exponent in this
formula and generalizing it to multiple levels.\,\cite{Blume68,Blume68b,Cianchi86}  

Here, we propose a simpler approach that has the advantage that it is
model-independent. 
To gain some insight we consider the limiting  case where
$H_{\rm rest}=0$ (equivalent to $\bb\Gamma=0$). By ensemble-averaging
over $\sigma_f=\pm 1$ we can write
\bean
G_d(\omega+i\eta)&=&\frac{1}{2}\frac{1}{\omega+i\eta-\omega_0+\lambda}+\frac{1}{2}\frac{1}{\omega+i\eta-\omega_0-\lambda}\nonumber\\
&=&\frac{1}{\omega+i\eta-\omega_0-\Sigma_d(\omega+i\eta)},\label{eq6}
\eean
where we have absorbed the statistical mixture into a perturbative
self-energy
$\Sigma_d(\omega+i\eta)=\lambda^2/(\omega+i\eta-\omega_0)$ for the ensemble-averaged Green's function. Note that
this agrees with Eq.\,\pref{eqExact} and Eq.\,\pref{eqDM}, while it
disagrees with Eq.\,\pref{eqGaussian} using $\chi_C(\tau)=-1$.  The structure of the $G_d$ function here 
is reminiscent of the zero-coupling limit of the Anderson impurity
problem. Indeed, for the case of $H_{\rm rest}=\sum_k\eps_kf\dg_k f\dn_k$
with a bandwidth $\Gamma$, we have an Ising Kondo (in a Zeeman field)
which can be solved exactly\,\cite{Komijani15} and $\chi_{-+}(\omega)$
exhibits orthogonality catastrophe physics. Usually, there is an
energy scale $T_*(\Gamma,\lambda)\sim \Gamma$, below which a coherent
entanglement between the probe and $H_{\rm rest}$ is established. However,
at the weak-coupling limit $T\gg T_*$ there is little entanglement and
a perturbative self-energy accurately matches the exact
non-perturbative result.\,\cite{Mao03,Komijani15}

One of observations of this paper is that the simplest ``single-photon''
exchange approximation to the self-energy of the $d$-fermion
provides an interpolation between motionally narrowed and double-line limits of the absorption line, and is capable of reproducing most of the observed
features of the lineshape for general $H_{\rm rest}$ in the $T\gg\Gamma$
limit.  This is the 
leading order contribution to the self
energy, $\Sigma_d(\tau)=-\lambda^2g_d(\tau)\chi_C(\tau)$, which to  order
$\lambda^2$ is exact and thus it goes beyond both forward-time
propagator and Gaussian-action approximations mentioned
earlier. Taking the Fourier transform, and using
$g_d(i\omega_n)=[i\omega_n-\omega_0]^{-1}$ and analytically continuing
to real frequencies we obtain (see app.\,5) \be
\Sigma''_d(\omega_0+\omega+i\eta)=\lambda^2[f(\omega_0)-n_B(-\omega)]\chi''_C(\omega+i\eta)\label{eqOurSelf}
\ee where $n_B(\omega)=[e^{\beta\omega}-1]^{-1}$ is the Bose-Einstein
distribution. Eq.\,\pref{eqOurSelf}
is the central result of this paper. We have related the retarded function of the fluctuations to the imaginary part of the self-energy. The latter is related to the absorbed power by $\Sigma_d''(\omega)=-cP(\omega)/[P^2(\omega)+ P'^2(\omega)]$ where $c$ is the proportionality constant introduced after Eq.\,\pref{eq3} and $P'(\omega)$ is the Hilbert transform of the absorbed power $P(\omega)\equiv P''$ so that $P'+i P''$ obeys the Kramers-Kr\"onig relation
\ben
P'(\omega)={\cal P}\int{\frac{d\omega'}{\pi}}\frac{P(\omega')}{\omega'-\omega}.\label{eqKK}
\een
Having obtained $\Sigma''_d$ one can then use Eq.\,\pref{eqOurSelf} to extract the charge susceptibility from the lineshape. A main assumption is that the fluctuations have a small bandwidth $\Gamma\ll T\ll\omega_0$. Thus, we can drop $f(\omega_0)$ and do a high-temperature expansion of $n_B(-\omega)\approx -1/\beta\omega$ to write $\Sigma''_d(\omega_0+\omega)\approx \lambda^2{\chi_C''/\beta\omega}$. This together with Kramers-Kr\"onig relation for $\chi_C(\omega+i\eta)$ provides another sum-rule that relates the area under the dissipative part of self-energy (or $\chi''_C/\beta\omega$) to the static charge susceptibility $\int{d\omega}\Sigma''_d(\omega)=\pi\lambda^2T\chi_C^0$. Moreover, combining these equations we find the simple and useful inverse formula
\be
{\chi_C(\omega)}\approx\frac{c}{\lambda^2T}\frac{\omega}{P'(\omega)+iP(\omega)}. 
\ee
This formula can be used to directly extract the charge susceptibility of the medium from the observed lineshape.

\begin{figure}[tp]
\includegraphics[width=1\linewidth]{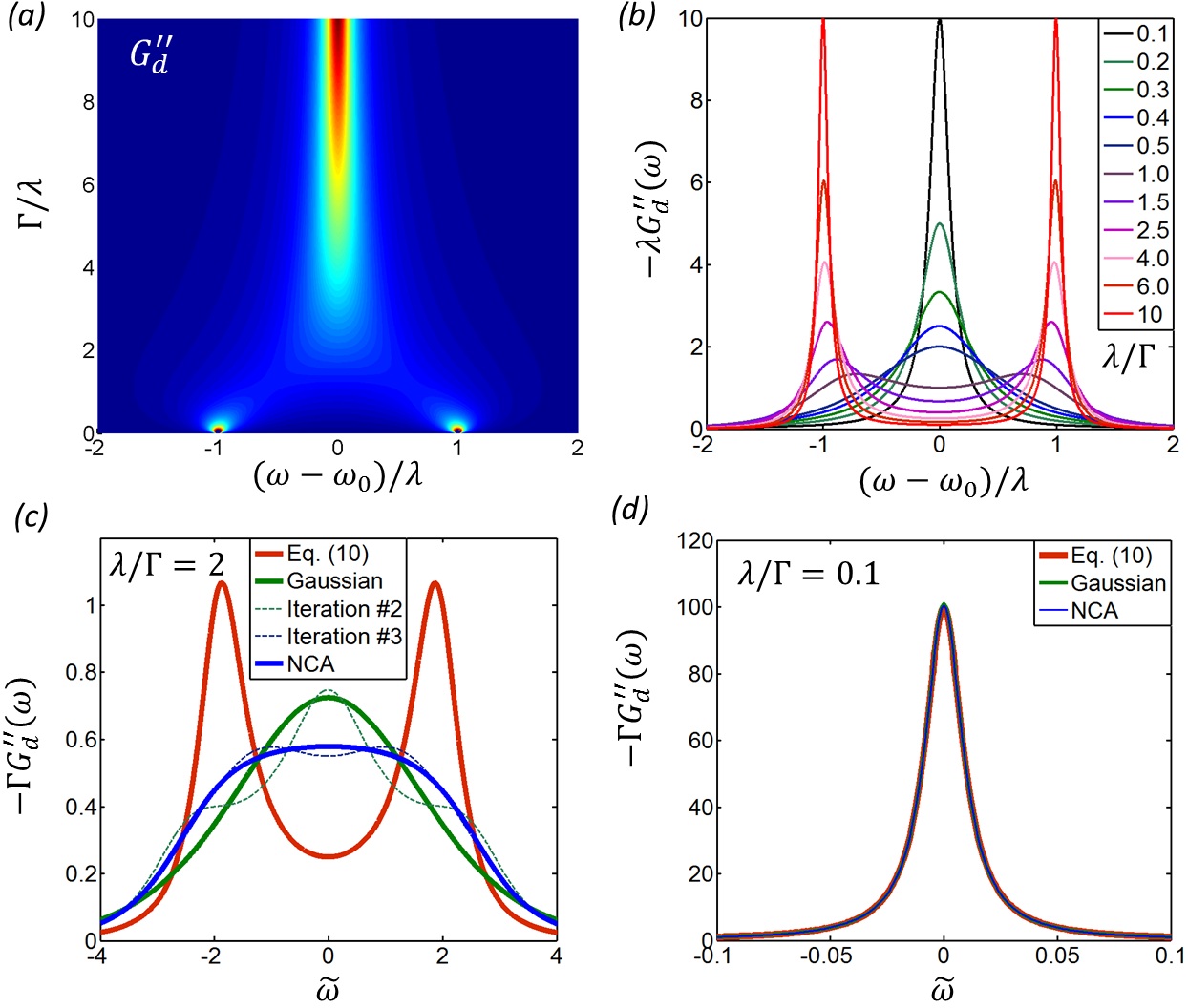}
\caption{\small \raggedright (color online) The resonance line-shape from a Lorentzian charge susceptibility from Eq.\,\pref{eqAA}. (a,b) For a fixed $\lambda$ varying $\Gamma$ leads to lineshape narrowing. (c) and (d) compares Eq.\,\pref{eqAA} with a (Markovian-) Gaussian calculation. Agreement at $\lambda/\Gamma\ll 1$ and disagreement at finite $\lambda/\Gamma$ is transparent. (c) also includes the non-crossing approxmation as well as the first two iterations of dressing the Green's function on Eq. (10), showing the splitting disappears and result becomes more and more similar to the Gaussian result.} \label{fig:Fig2}
\end{figure}

\section{Examples}

As a first example, we look at the case when $\chi''_C/\beta\omega=-\pi\delta_\Gamma(\omega)$ is a $\Gamma$-broadened delta function at the origin. A Lorentzian $\delta
_\Gamma(\omega)$ function, defined in Eq.\,\pref{eq60}, corresponds to an exponential decay in the time-domain $\braket{\sigma_f(t)\sigma_f(0)}\sim e^{-\Gamma t}$ and a single-rate Markovian process in which $\Gamma$ is the rate of switching between the two $\sigma_f=\pm 1$ states (see app.\,4).
In this case, Kramers-Kr\"onig relation is trivial and we obtain $\Sigma_d(\omega_0+\omega+i\eta)=\lambda^2[\omega+i\Gamma]^{-1}$. Inserting this into the Green's function 
gives us the \emph{Archer-Anderson} formula:
\be
G''_d(\omega_0+\omega+i\eta)=-\frac{\lambda^2\Gamma}{(\omega^2-\lambda^2)^2+\omega^2\Gamma^2},\label{eqAA}
\ee
previously\,\cite{Anderson54} obtained from the completely different approach of Eq.\,\pref{eqDM} (see app.\,4). This agreement is a remarkable observation whose origin is unclear to us at the moment. 
Eq.\,\pref{eqAA} is plotted in Fig.\,(3 a,b).
In the slow-switching case $x\equiv\lambda/\Gamma\gg 1$, we have two well separated $\delta$-peaks at $\omega=\pm\lambda$, whereas in the fast-switching case $x\ll 1$, and on a rescaled frequency axis $\tilde\omega=(\omega-\omega_0)/\Gamma$, we get a single peak at the origin $\Gamma G''_d(\tilde\omega)\to x^2[\tilde\omega^4+\tilde\omega^2+x^4]^{-1}$ which is Lorentzian within $\tilde\omega\ll x$ and has non-Lorentzian $\tilde\omega^{-4}$ tails at $\tilde\omega\gg x$. 

It is instructive to compare Eq.\,\pref{eqAA} to the Gaussian result, Eq.\,\pref{eqGaussian} which can be computed with little effort in this special case. Defining $\alpha=x^2$ and from $\chi(\tau)=-e^{i\Gamma\tau}$, it is straightforward (app.\,4) to find
\be
\Gamma G''_d(\tilde\omega)=-\pi e^{\alpha}\sum_{n\ge 0} f_n(\alpha)\delta_{(n+\alpha)}(\tilde\omega).\label{eqMG}
\ee
This is a summation of $(n+\alpha)$-broadened $\delta$-functions all centred at $\tilde\omega=0$ with the coefficients $f_n(\alpha)\equiv\sum_mJ_{m}(-\alpha)I_{n-m}(-\alpha)$, given in terms of ordinary $J_m(x)$ and modified $I_m(x)$ Bessel functions. This function is plotted in Fig. (3c) and (3d) along with Eq.\,\pref{eqAA}. The two functions agree at $x\ll 1$, which can be understood, qualitatively, by the fact that the time-averaged $\overline{\sigma_f}$ in both continuous and discrete cases, spend most of the time around zero. We argue below that the rainbow diagrams [e.g. second diagram of Fig.\,(1b)] and vertex corrections [e.g. third diagram of Fig.\,(1b)] that are included in the Gaussian subset but not in Eq.\,\pref{eqOurSelf}, are negligible for $x\ll 1$ but become important at finite $x$ in agreement with these plots.

\begin{figure}[tp]
\includegraphics[width=1.0\linewidth]{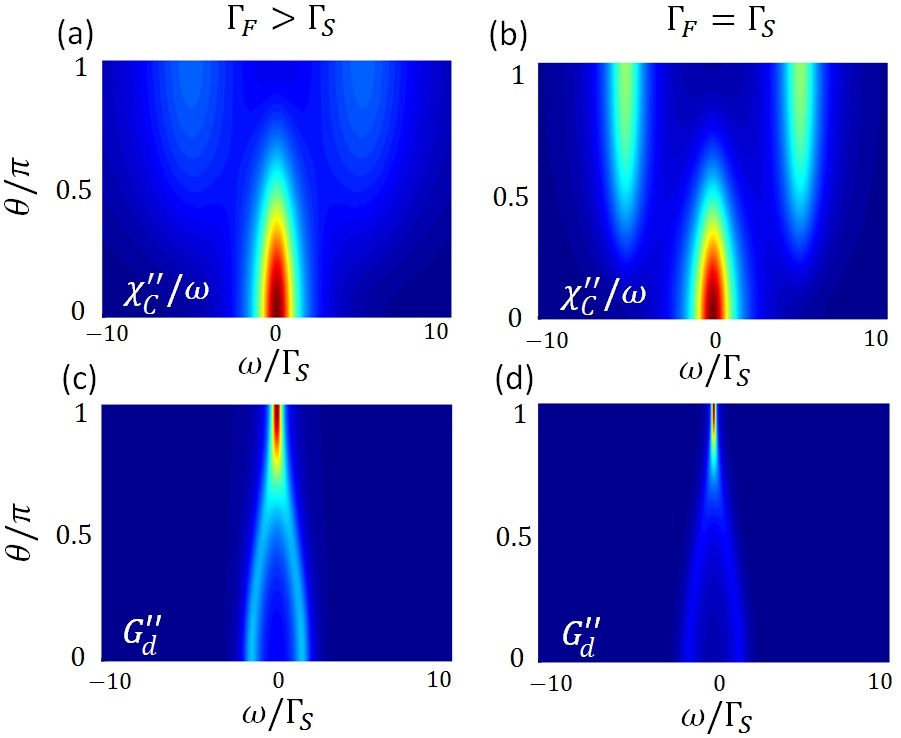}
\caption{\small\raggedright (color online) The effect of non-trivial charge fluctuations on the resonance lineshape where there are two (slow with the width $\Gamma_S$ and fast with the width $\Gamma_F$) contributions to the charge susceptibility. (a) and (b) correspond to the charge susceptibility.
We have varied the relative strength of the two proportional to $\sin^2\theta$ and $\cos^2\theta$, and calculated the resulting line shape (c) and (d). 
}
\end{figure}

\gray{Fig.\,\pref{fig:Fig1} puts the retained diagrams in perspective with the other possible Feynman diagrams. In comparison to Eq.\,\pref{eqGaussian}, besides dropping the connected diagrams, we have additionally dropped the rainbow diagrams [e.g. Fig.\,(1f)] as well as crossing diagrams [e.g. Fig.\,(1g)] which represent vertex corrections to our leading-order self-energy. In the following we show that these corrections are negligible in the limit of fast charge fluctuations, but they become important in the opposite slow regime.} 

To account for the vertex correction one has to solve the Bethe-Salpeter equation for the vertex function (see app.\,6). To the leading order, and from Fig.\,(2d), we can estimate these vertex corrections to be $\lambda^2\to \kappa\lambda^2$ where $\kappa\approx 1-\lambda^2\int_0^{\tau}d\tau_1\chi_C(\tau_1)\int_0^{\tau_1}d\tau_2\approx 1-x^2$. This is analogous to the Migdal's theorem in superconductivity and confirms that vertex corrections are negligible in the fast regime $x\ll 1$ but important in the slow regime. 

The rainbow diagrams are part of the so-called non-crossing approximation (NCA) class and can be included within our formalism by self-consistently upgrading the bare propagator $g_d$ in the self-energy to the exact Green's function $G_d$. This is schematically shown in Fig.\,(2c) and can be implemented iteratively. Figs.\,(3 c,d) include the first two iterations as well as the NCA result, starting from Eq.\,\pref{eqAA}. While nothing changes for $x\ll 1$, in the opposite regime of finite $x$ the lineshape becomes closer to the Gaussian result. To the leading order we can see this, by expanding the self-energy as $\Sigma_d(\omega)=\Sigma_d(\omega_0)+(\omega-\omega_0)\partial_\omega\Sigma_d(\omega_0)+\cdots$ and only keeping the leading terms. Therefore, $g_d(i\omega_n)$ is replaced by $G_d(i\omega_n)\approx Z[i\omega_n-\omega_0+i\Lambda\sgn(n)]^{-1}$ where $Z=[1-\partial_\omega\Sigma_d(\omega_0)]^{-1}$ is the wavefunction renormalization of the $d$-level and $\Lambda=-Z\Sigma''_d(\omega_0)$ is its broadening.
\gray{, i.e. the natural linewidth of the two-level system as a result of its coupling to the environment}
We can estimate $\Lambda/\Gamma\propto x^2$ and $Z=[1-x^2]^{-1}$ in our example. 
Calculating the self-energy with this propagator has the effect of renormalizing the coupling constant $\lambda^2\to \lambda^2Z$, but more importantly the self-energy is blurred by a convolution with a $\Lambda$-broadened delta function $\delta_\Lambda(\omega)$ (see app.\,5). Therefore, in the limit of fast/slow charge fluctuations rainbow diagrams have a negligible/important effect, in agreement with the numerical result. The fact that rainbow diagrams and vertex corrections drive the lineshape towards the Gaussian result is not surprising, since these additional diagrams can be understood as if the two-level system is coupled to many independent mutually-non-interacting $f$-levels which average into a Gaussian lineshape.

The second example is when a combination of fast and slow switching processes are involved and where the power of the field theoretical approach shows up. These more complex scenarios happen for example in Kondo systems as spin-flip processes involve fast charge switching between unpaired single-spin and empty or doubly occupied states.\,\cite{Piers} In Fig. (3), we have considered a case where $\chi''_C/\beta\omega$ contains a slow peak at the origin with the width $\Gamma_S$ and a fast mode at a finite frequency with the width $\Gamma_F$. The charge susceptibility is then
\be
\frac{\chi''_C(\omega)}{\beta\omega}=-\pi\cos^2\theta\delta_{\Gamma_S}(\omega)-\pi\sin^2\theta\delta_{\Gamma_F}(\abs{\omega}-\omega_F).
\ee
where $\omega_F$, $\Gamma_S$ and $\Gamma_F$ are kept constant and $\theta$ is varied in Figs.\,(3a,b) to change the relative strength of fast and slow fluctuations for different ratios of $\Gamma_F<\Gamma_S$ and $\Gamma_F=\Gamma_S$. 
Figs.\,(3c,d) shows the corresponding lineshapes we would expect from such a medium.
A direct consequence of the sum rules is that the total area under $\chi''_C/\omega$ is a constant. As is clear from these results, the existence of the fast mode and its breadth can have significant effects on the lineshape, a detailed study of which we leave for the future.\\

\section{Conclusion}

To conclude, we have revisited the classic stochastic theory of lineshapes and used a Majorana representation of the spin to recast it in a field-theoretical language. We have shown that the leading contribution to the self-energy provides an interpolation between Markovian and Gaussian results as well as featuring a flexibility to study non-trivial charge fluctuations. \\

\section*{Acknowledgement}
 
We would like to thank H.~Kobayashi and S.~Nakatsuji for valuable
discussions and communicating their measurement results before
publication.  This work was supported by NSF grant DMR-1309929 (Piers Coleman)
and a Rutgers University Materials Theory postdoctoral fellowhsip
(Yashar Komijani).


\section{Appendix: Multi-level systems}
The ideas discussed in the paper can be easily generalized to multi-level system. In the following, we discuss a three-level system, but generalization to more levels is straightforward. We model the three levels $\ket{g}$, $\ket{e_1}$ and $\ket{e_2}$ by two spins $S_1$ and $S_2$ (two two-level systems) with the additional constraint that forbids them to be both in their excited states at the same time. The Hamiltonian is
\bea
H&=&(\omega_1+\lambda_1\sigma_f)I^z_1+(\omega_1+\lambda_2\sigma_f)I^z_2\\
&&\qquad+U(I^z_1+1/2)(I^z_2+1/2)+H_{\rm rest}
\eea
and assume that $U\to\infty$. Following similar discussions as in the two-level system by including a photonic degree of freedom so that
\be
H'=H+g_1(I^+_1a+h.c.)+g_2(I^+_2a+h.c.)+\omega a\dg a
\ee
for the absorbed power we obtain
\be
P=\omega\Omega^2\{g_1^2\chi''_{11}(\omega)+g_2^2\chi''_{22}(\omega)+g_1g_2[\chi''_{12}(\omega)+\chi''_{21}(\omega)]\}
\ee
Here, $\chi''_{ij}(\omega)$ is the Fourier transform of the retarded function $-i\theta(t)\braket{\com{I^-_i(t)}{I^+_j(0)}}$. The last two terms are coherent factors that when vanish, the spectra represents two independent two-level systems. For each two-level system, we use Majorana representation and introduce a fermionic $d$-level so that in analogy with the paper
\be
P\propto\{g_1^2G''_{11}(\omega)+g_2^2G''_{22}(\omega)+g_1g_2[G''_{12}(\omega)+G''_{21}(\omega)]\}
\ee
\end{comment}


\appendix
\section*{Appendix}
The following appendices include the proof of some of the statements or formulas given in the manuscript.

\subsection{Absorbed power formula}
From Fermi's Golden rule, the absorption rate 
is related to the transition rate, given by
\be
\Gamma(\omega)=\sum_f\frac{2\pi}{\hbar}\abs{\braket{f\vert \delta H\vert i}}^2\delta(E_f-E_i)
\ee
Here $\delta H=g(a\dg I^-+h.c.)$ and $E_i$ and $E_f$ are the exact many-body energies of $H_0=H+\omega a\dg a$ where $H$ is the Hamiltonian in Eq.\,\pref{eq1}. Then using
\be
\delta(E_f-E_i)=\frac{1}{2\pi \hbar }\intinf{dt}e^{-it(E_f-E_i)/\hbar }
\ee
we can write 
\begin{eqnarray}\label{l}
\Gamma(\omega)
&=&
\frac{1}{\hbar^{2}}\int_{-\infty }^{\infty }dt e^{-it (E_f-E_{i} ) /\hbar }
\sum_f\bra{i}
 \delta H 
\ket{f}\bra{f}\delta H\ket{i}\cr
&=&
\frac{1}{\hbar^{2}}\int_{-\infty }^{\infty }dt
\sum_f\bra{i} 
e^{itH_{0}/\hbar } \delta H e^{-itH_{0}/\hbar }
\ket{f}\bra{f}\delta H\ket{i}\cr
&=& \frac{1}{\hbar^{2}}\int_{-\infty }^{\infty }dt
\bra{i} \delta H (t)\delta H (0)
\ket{i},
\end{eqnarray}
where we have employed completeness $\sum_{f}\vert f \rangle \langle
f \vert = 1 $ and written  the perturbation
\[
\delta H (t) = e^{itH_{0}/\hbar } \delta H e^{-itH_{0}/\hbar }
\]
in the interaction representation with respect to $H_0=H+\omega a\dg a$. As a result of this interaction representation $I^\pm$, $a$ and $a\dg$ in $\delta H$ develop time-dependence.
Doing a thermal average over the initial states, the total transition
rate is then given by [the frequency dependence on the right is implicit in the time-evolution of the $\delta H(t)$].
\be
\Gamma(\omega)=\frac{1}{\hbar^{2}}\intinf{dt\braket{\delta H(t)\delta H(0)}}
\ee
which can be written as
\bea
\Gamma(\omega)&=&
\left(\frac{g}{\hbar } \right)^{2}
\intinf{dt}\{ne^{i\omega t}\braket{I^-(t)I^+(0)}\nonumber\\
&&\hspace{2cm}+(1+n)e^{-i\omega t}\braket{I^+(t)I^-(0)}\}\nonumber\\
&=&ig^2[n\chi_{-+}^>(\omega)+(1+n)\chi_{-+}^<(\omega)]\qquad\label{eq15}
\eea
where $n\equiv\braket{a\dg a}$ as the average number of photons created at the steady state (a measure of input power). We have used the definitions of the greater/lesser functions
\be
\chi_{-+}^>(t)=-i\braket{I^-(t)I^+(0)}, \quad
\chi_{-+}^<(t)=-i\braket{I^+(0)I^-(t)}.\nonumber
\ee
We assume in this paper that the two-level system (described by the spin $\vec I$ and weakly probed by the gamma photons) is in thermal equilibrium with its surrounding and considering the $\omega_0\gg T$ condition, it is mainly in its ground state. Therefore, the greater/lesser susceptibilities are related to the imaginary part of retarded susceptibility by the fluctuation-dissipation theorem
\bea
\chi_{-+}^>(\omega)&=&2i[1+n_B(\omega)]\chi''_{-+}(\omega),\nonumber\\
\chi_{-+}^<(\omega)&=&2in_B(\omega)\chi''_{-+}(\omega)\label{eqequil}
\eea
Anticipating that $\chi''_{-+}(\omega)$ is a narrow resonance at $\omega_0\gg T$, we can approximate the Bose function $n_B(\omega)=[e^{\beta\omega}+1]^{-1}\approx\theta(-\omega)$. Therefore, $\chi^<_{-+}(\omega\gg T)\approx 0$ and $\chi^>_{-+}(\omega\gg T)\approx 2i\chi''_{-+}(\omega)$. Inserting these into Eq.\,\pref{eq15} we conclude
\be
\Gamma(\omega)=-2\Omega^2\im{\chi_{-+}^R(\omega)}\label{eq16}
\ee
where we have defined $\Omega\equiv g\sqrt{n}$. Eq.\,\pref{eq16} is the total number of transitions.  Dividing this by two (absorptions) and writing $P(\omega)=\frac{1}{2}\hbar\omega\Gamma(\omega)$ gives Eq.\,\pref{eq2}.
\subsection{Majorana representation of the spin}
\subsubsection{Commutation relations}
Using the index notation $S^l=- (i/2)\eps_{lab}\eta^a\eta^b$, 
and using the anticommutation algebra
$\{\eta_{a},\eta_{b} \}= \delta_{ab}$, 
we can confirm that the spin operators faithfully reproduce the 
SU(2) algebra
\begin{eqnarray}
[S^l,S^m]&=&
-\frac{1}{4}\eps_{lab}\eps_{mcd}[\eta_{a}\eta_{b},\eta_{c}\eta_{d}]\cr
&=& 
-\frac{1}{4}\eps_{lab}\eps_{mcd}\Big\{
\delta^{bc} \eta^a\eta^d -\delta^{bd}\eta^a\eta^c\nonumber\\
&&\hspace{2cm}
+\delta^{ac}\eta^d\eta^b -\delta^{ad}\eta^c \eta^b\Big\}\cr
&=& \frac{1}{2}\left[\eta^{l}\eta^{m} -\eta^{m}\eta^{l}\right] = i \epsilon_{lmp}S^{p}.
\end{eqnarray}
and that furthermore, $(S^{a})^{2}=1/4$, $\vec{S}^{2}=\frac{3}{4}$,
confirming that this is a faithful representation of a spin-1/2 operator. 

\subsubsection{Relation between Keldysh and retarded functions}
As a reminder, we can write the Kelydsh $G^K$ and the difference between retarded $G^R$ and advanced $G^A$ functions as 
\be
G^K_d=G^>+G^<,  \qquad G^>-G^<=G^R-G^A,\label{eqGdef}
\ee
in terms of the greater and lesser functions, defined by
\be
G^>_d(t)=-i\braket{d(t)d\dg(0)}, \quad
G^<_d(t)=+i\braket{d\dg(0) d(t)}.\nonumber
\ee
Using the cyclic properties of the trace, $G_d^>(t-i\beta)=-G_d^<(t)$  or equivalently, $G^<_d(\omega)=-e^{-\beta\omega}G^>_d(\omega)$ in the frequency domain. Combining these with Eqs.\,\pref{eqGdef} gives $G^K_d(\omega)=[1-2f(\omega)]2iG''_d(\omega)$ which leads to Eq.\,\pref{eq3}.

\subsection{Diagrammatic proofs}
The Hamiltonian \pref{eq1} in the $d$-level representation is
\be
H=(\omega_0+\lambda\sigma_f)d\dg d+H_{\rm rest}\{\sigma_f\}\label{eq33}
\ee
and we are interested in $G_d(\tau)=-\braket{T_\tau d(\tau)d\dg(0)}$. A Brute force perturbation theory in $\lambda$ is
\bea
G_d(\tau)&=&-\sum_{n=0}^{\infty}\frac{(-\lambda)^n}{n!}\intob{d\tau_1\cdots d\tau_n}\nonumber\\
&&\hspace{-1.2cm}\braket{T_\tau d(\tau)d\dg(0)[d\dg(\tau_1)d(\tau_1)\sigma_f(\tau_1)]\dots [d\dg(\tau_n)d(\tau_n)\sigma_f(\tau_n)]}.\nonumber
\eea
The $d$ operaotrs commutes with $H_{\rm rest}\{\sigma_f\}$ and therefore, there is no additional interaction vertex and the above correlation function factorizes into $d$ and $\sigma_f$ parts. Therefore, we can write
\bea
G_d(\tau)&=&-\sum_{n=0}^{\infty}\frac{(-\lambda)^{n}}{n!}\intob{d\tau_1\dots d\tau_{n}}\nonumber\\
&&\hspace{-1cm}\braket{T_{\tau}d(\tau)d\dg(0)d\dg(\tau_1)d(\tau_1)\cdots d\dg(\tau_n)d(\tau_{n})}\times\nonumber\\
&&\hspace{3.5cm}
\braket{T_{\tau}\sigma_f({\tau_1})\cdots \sigma_f({\tau_n})}.\qquad
\eea
We can apply Wick's contraction to these non-interacting $d$-levels, but before that we introduce an approximation that simplifies the resulting diagrams. 

\subsubsection{Forward-time propagators and single-branch simplification}
The propagator for the free $d$-level is
\bea
g_d(\tau)&\equiv&-\braket{T_\tau d(\tau)d\dg(0)}\nonumber\\
&=&\{-\theta(\tau)[1-f(\omega_0)]+\theta(-\tau)f(\omega_0)\}e^{-\omega_0\tau}\nonumber\\
&=&[f(\omega_0)-\theta(\tau)]e^{-\omega_0\tau}\approx -\theta(\tau)e^{-\omega_0\tau}.\label{eqFt}
\eea
We have used that the (average) resonant frequency is much larger than temperature ($\omega_0\gg T$) and therefore
\be
f(\omega_0)=\frac{1}{e^{\beta\omega_0}+1}\approx e^{-\omega_0/T}\approx 0.
\ee

In the following, we show that this approximation significantly simplifies the wick's contractions. We can categorize the Feynman diagrams with the number of inter-disconnected $d$-level propagators. If going from $d(0)$ to $d(\tau)$ we pass through all the $d$-level propagators, there is only one branch. Inter-disconnected diagrams may appear for example to order $\lambda^2$ as the tadpole diagram or $\lambda^4$ as the fermion bubbles inside the propagator of the $\sigma_f$ charge fluctuation [Fig.\,(2b)]. 
The simplification of the forward-time approximation is that, $d$-fermion bubbles are suppressed. For example, the fermion bubbles' contribution is proportional to $g(\tau_1-\tau_2)g(\tau_2-\tau_1)\approx\theta(\tau_1-\tau_2)\theta(\tau_2-\tau_1)=0$. Therefore, single-branch assumption is justified. 

Moreover, since $g_d(\tau<0)=0$, in doing Wick's contraction, we must have the ascending order $\tau_1<\tau_2<\cdots<\tau_n$ and they are all connected in one branch. If the order is violated even once, e.g. $g(\tau_1-\tau_3)$ the result will be zero because $\tau_2$ will appear somewhere else and make the time-argument of the propagator negative and zero result due to forward-time approximation. With this, we have
\bea
G_d(\tau)&=&\sum_{n=0}^{\infty}{\lambda^{n}}\int_0^{\tau}{d\tau_1}
\int_0^{\tau_1}{d\tau_2}
\cdots
\int_0^{\tau_{n-1}}{d\tau_n}\nonumber\\
&&
g(\tau-\tau_1)g(\tau_1-\tau_2)
\cdots
g(\tau_{n-1}-\tau_n)
g(\tau_n-0)\times\nonumber\\
&&\hspace{3.8cm}
\braket{\sigma_f({\tau_1})\cdots \sigma_f({\tau_n})}.\qquad
\eea
Using Eq.\,\pref{eqFt} we find
\bea
G_d(\tau)&=&-e^{-\omega_0\tau}\sum_{n=0}^{\infty}{(-\lambda)^{n}}\int_0^{\tau}{d\tau_1}
\int_0^{\tau_1}{d\tau_2}
\cdots
\int_0^{\tau_{n-1}}{d\tau_n}\nonumber\\
&&\hspace{4.5cm}
\braket{\sigma_f({\tau_1})\cdots \sigma_f({\tau_n})}\nonumber\\
&=&-e^{-\omega_0\tau}\sum_{n=0}^{\infty}\frac{(-\lambda)^{n}}{n!}\int_0^{\tau}{d\tau_1}
\int_0^{\tau}{d\tau_2}
\cdots
\int_0^{\tau}{d\tau_n}\nonumber\\
&&\hspace{4cm}
\braket{T_\tau\sigma_f({\tau_1})\cdots \sigma_f({\tau_n})}\nonumber\\
&=&-e^{-\omega_0\tau}\braket{T_\tau e^{-\lambda\int_0^{\tau}d\tau'\sigma_f(\tau')}}.\label{eq42}
\eea
which is Eq.\,\pref{eqExact}.
\subsubsection{Orthogonality catastrophe proof}
An alternative proof of Eq.\,\pref{eq42} is by writing the evolution operator of the $d(\tau)=e^{\tau H}de^{-\tau H}$, where $H$ is given by Eq.\,\pref{eq33}. To have non-zero result, the $d$-state has to be filled  prior to the acting with the annihilation operator and empty afterwards. Therefore, we can write
\be
d(\tau)=e^{\tau H_{\rm rest}}de^{-\tau [H_{\rm rest}+(\omega_0+\lambda\sigma_f)]}.
\ee
After this replacement, the $d$ operator has done its job and can be dropped out\,\cite{Komijani15} :
\bea
G_d(\tau)&=&-\braket{T_\tau d(\tau)d\dg}\nonumber\\
&\approx&-\braket{e^{\tau H_{\rm rest}}e^{-\tau [H_{\rm rest}+(\omega_0+\lambda\sigma_f)]}}_{H_{\rm rest}}\nonumber\\
&=&-e^{-\omega_0\tau}
\braket{e^{\tau H_{\rm rest}}e^{-\tau (H_{\rm rest}+\lambda\sigma_f)}}_{H_{\rm rest}}
.\label{eq45}
\eea
We have added a subindex $H_{\rm rest}$ to the first correlator to indicate the Hamiltonian that appears in the corresponding Boltzman factor. Also, we have used the forward-time approximation to drop a factor proportional to $e^{-\beta\omega_0}$ in the expansion of the partition function
\bea
Z&=&\tr{e^{-\beta H_{\rm rest}}}+e^{-\beta\omega_0}\tr{e^{-\beta[H_{\rm rest}+\lambda\sigma_f]}}\nonumber\\
&\approx& \tr{e^{-\beta H_{\rm rest}}}.
\eea
The product of the two exponents $U(\tau)\equiv e^{\tau H_{\rm rest}}e^{-\tau(H_{\rm rest}+\lambda\sigma_f)}$ in Eq.\,\pref{eq45} is the definition of the time-evolution in the interaction picture with respect to $H_{\rm rest}$. To see that, just take its derivative with respect to $\tau$ and observe the corresponding Schr\"odinger equation that it obeys. 
\bea
\frac{d}{d\tau}U(\tau)&=&e^{\tau H_{\rm rest}}[H_{\rm rest}-(H_{\rm rest}+\lambda\sigma_f)]e^{-\tau(H_{\rm rest+\lambda\sigma_f})}\nonumber\\
&=&-\lambda\hat\sigma_f(\tau)U(\tau)
\eea
where $\hat\sigma_f(\tau)=e^{\tau H_{\rm rest}}\sigma_fe^{-\tau H_{\rm rest}}$. Integrating this equation, we can write
\be
G_d(\tau)\approx -e^{-\omega_0\tau}\braket{T_\tau e^{-\lambda\int_0^\tau d\tau'\sigma_f(\tau')}}.
\ee
\subsubsection{Gaussian subset}
Generally, the $n$-point function of $\sigma_f$-s can all have non-trivial connected and disconnected ones. If $H_{\rm rest}$ is Gaussian, all the connected terms vanish. The set of all disconnected terms [exact for Gaussian $H_{\rm rest}$], e.g. the diagram shown in Fig.\,(5)a can be written as Anderson wrote
\bea
G_d^D(\tau)&=&-e^{-\omega_0\tau}\exp\Big[{\frac{\lambda^2}{2}\int_0^\tau \int_0^\tau{d\tau_1}{d\tau_2}\braket{T_\tau \sigma_f(\tau_1)\sigma_f(\tau_2)}}\Big]\nonumber\\
&=&e^{-\omega_0\tau}\exp\Big[{-\lambda^2\int_0^{\tau}(\tau-x)\chi_C(x)dx}\Big].
\eea
This contains all order of $\lambda^{2n}$, each being sum of all possible two-point contractions of $2n$ $\sigma_f$-operators. Some of these are shown in the following figure.

\begin{figure}[h!]
\includegraphics[width=.7\linewidth]{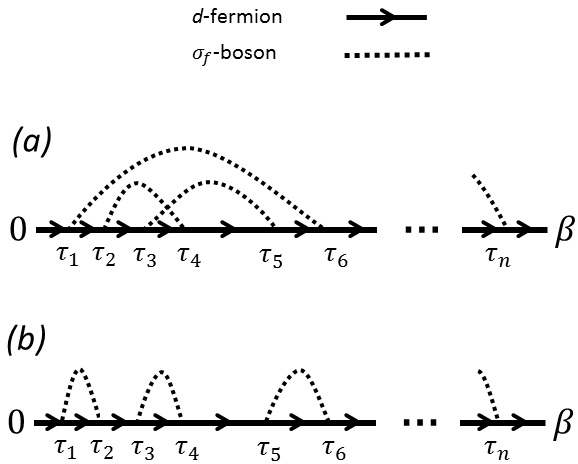}
\caption{\small (a) An example of an $O(\lambda^n)$ diagram within Gaussian (and forward-time propagator) approximation. (b) The diagrams obtained from the leading self-energy contribution. }
\end{figure}

\subsection{Markovian case}
\subsubsection{Proof of Eq.\,\pref{eqDM}}
In the Markovian case, the main assumption is that the reduced density matrix of a single $\sigma_f$ is diagonal and it obeys the classical rate equation. Then Eq. (4) has to be interpreted as a statistical average. Following Anderson we divide the interval $(0,\tau)$ to $N$ segments and write Eq. (4) as
\bea
&&\braket{T_\tau e^{\lambda\int_0^\tau d\tau'\sigma_f(\tau')}}=\braket{\exp\Big[\lambda\frac{\tau}{N}\sum_{m=1}^N\sigma_{f,m}\Big]}\nonumber\\
&&=\sum_{\sigma_{f,1}=\pm1}\dots\sum_{\sigma_{f,N}=\pm1}
p(\sigma_{f,N};\sigma_{f,N-1};\dots;\sigma_{f,1})\times\nonumber\\
&&\hspace{5cm}
\prod_{m=1}^N \exp\Big[{\lambda\frac{\tau}{N}\sigma_{f,m}}\Big]\qquad.\label{eq52}
\eea
$p(\sigma_{f,N};\sigma_{f,N-1};\dots;\sigma_{f,1})$ is the probability that the variable $\sigma_f$ is equal to $\sigma_{f,1}$ at time $\tau/N$, $\sigma_{f,2}$ at time $2\tau/N$, $\sigma_{f,n}$ at time $n\tau/N$ and so on. The Markovian assumption means that the probability of the state $\sigma_{f,n}$ only depends on the state $\sigma_{f,n-1}$ and hence
\bea
&&p(\sigma_{f,N};\sigma_{f,N-1};\dots;\sigma_{f,1})=p(\sigma_{f,N}\vert\sigma_{f,N-1},\tau/N)\times\nonumber\\
&&\hspace{4.2cm}
p(\sigma_{f,N-1}\vert\sigma_{f,N-2},\tau/N)\times\nonumber\\
&&\hspace{5.6cm}\dots\nonumber\\
&&\hspace{4.2cm}
p(\sigma_{f2}\vert\sigma_{f1},\tau/N)p(\sigma_{f1}).\label{eq53}
\eea
Using the (imaginary time) rate equation $d\vec p/d\tau=-i\bb\Gamma \vec p$, one can write
\be
p(\sigma_{f,m}\vert\sigma_{f,m-1},\tau/N)=\bb 1-i\frac{\tau}{N}\bb\Gamma\approx \exp[-i\frac{\tau}{N}\bb\Gamma].
\ee
This can be combined with Eq.\,\pref{eq53} to write Eq.\,\pref{eq52} as a product of matrices
\bea
\braket{T_\tau e^{\lambda\int_0^\tau d\tau'\sigma_f(\tau')}}&=& \vec 1\hphantom{.} ^T\cdot\Big[\exp[-i\frac{\tau \bb\Gamma}{N}]\exp[\lambda\frac{\tau \sigma^z}{N}]\Big]^{N-1}\vec p_1,\nonumber
\eea
where $\sigma^z$ is a Pauli matrix containing the two possible states of each time-slot. We can combine the two matrices and drop the commutator term of order $1/N^2$. Replacing $\vec p_1$ with the steady-state probabilities and using $N\to\infty$ limit we get the desired result
\be
G_d(\tau)\to-e^{-\omega_0\tau} \{\vec 1\hphantom{ .}^T.\exp[\tau(\lambda\sigma^z-i\bb\Gamma)].\vec p_{st}\}.\label{eq49}
\ee
\subsubsection{Archer-Anderson formula}
To calculate the Green's function from this method, one has to diagonalize the matrix $\bb M=\lambda\sigma^z-i\bb\Gamma$ in the exponent and expand the $\vec p_{st}$ in terms of the eigenstates of that matrix. For a single-rate Markovian model,
\be
\bb\Gamma=\frac{\Gamma}2\mat{-1 & 1 \\ 1 & -1},
\ee
from which using $\bb\Gamma\vec p_{st}=0$ we have
\be
\bb M=\frac{\Gamma}{2}\mat{2x+i & -i \\ -i & -2x+i}, \qquad
\vec p_{st}=\frac{1}{2}\mat{1\\ 1}
\ee
where $x=\lambda/\Gamma$ as before, and the (right-) eigenvalue/vectors of $\bb M$, defined by $\bb M \vec u_\pm=v_\pm \vec u_\pm$, are
\be
v_\pm=\frac{\Gamma}{2}[i\mp\sqrt{4x^2-1}], \qquad \vec u_{\pm}=\frac{1}{\sqrt 2}\mat{i\sqrt{1\mp y}\\ \sqrt{1\pm y}}\nonumber
\ee
with $y\equiv\sqrt{1-1/4x^2}$. In the slow limit $x\gg 1$ the first term of the eigenvalue is the width and the second term is the splitting, but in the fast limit $x\ll 1$ they merge into one peak. Substituting in Eq.\,\pref{eq49} we get
\bea
G_d(\tau)=-e^{(-\omega_0+i\Gamma/2)\tau}&\Big[&q_+'{q\dn_+}e^{-(\tau\Gamma/2)\sqrt{4x^2-1}}\nonumber\\
&&\hspace{1cm}+q_-'{q\dn_-}e^{(\tau\Gamma/2)\sqrt{4x^2-1}}\Big].\qquad\label{eq61}
\eea
Here, $q_{\pm}$ are the coefficients of the expansion $\vec p_{st}=q_+\vec u_++q_-\vec u_-$, given by
\be
q_\pm=\frac{\pm 1}{2\sqrt 2 y}[\sqrt{1\pm y}+i\sqrt{1\mp y}],
\ee
and $q_\pm'=\vec 1^T\cdot \vec u_\pm$ are related to them by $q'_\pm=\pm 2yq_\pm$. Substituting these into Eq.\,\pref{eq61}, Fourier transforming, and doing the analytical continuation gives
\bea
G_d(\omega+i\eta)&=&\frac{\omega+i\Gamma}{
\omega(\omega+i\Gamma)-x^2\Gamma^2}
=\frac{1}{\omega-\lambda^2/(\omega+i\Gamma)},\qquad
\eea
in agreement with our self-energy result. The imaginary part gives the Archer-Anderson formula, Eq.\,\pref{eqAA}. Note that we assumed that $x>1$ in our proof. Interestingly, exactly the same result is obtain in the opposite regime of $x<1$. While this comes as a surprise\,\cite{Anderson54} in the the present derivation, our self-energy derivation in the paper makes it clear that indeed there is no analytical difference between these two limits.

\subsubsection{Markovian charge susceptibility}
We can obtain this from the rate equation for the probabilities $p_{\pm}$ of being in the states $\sigma_f=\pm 1$, i.e. $\dot p_{\pm}=\Gamma_{\pm}p_{\mp}-\Gamma_{\mp}p_{\pm}$. Writing $\Gamma_{\pm}=\frac{1}{2}\Gamma(1\pm\braket{\sigma_f}_{s})$, and combining the two equations to describe the dynamics of 
 $\braket{\sigma_f}=p_+-p_-$ we obtain the Bloch equation
\be
\frac{d}{dt}\braket{\sigma_f}=-\Gamma[\braket{\sigma_f}-\braket{\sigma_f}_{s}]\label{eq9}
\ee
where $\braket{\sigma_f}_s=\Gamma_+-\Gamma_-$ is the steady-state value given by the mismatch in tunnelling rates. A small slowly-varying polarizing field $h(t)\sigma_f$ changes the probabilities to $p_\pm\to e^{\mp\beta h(t)}/(e^{\beta h(t)}+e^{-\beta h(t)})\approx [1\mp\beta h(t)]/2$ and leads to the steady-state value $\braket{\sigma_f}_s=\chi_0 h(t)$ with $\chi_0=-\beta$. Therefore, taking the Fourier transform of Eq.\,\pref{eq9}, the susceptibility can be obtained from the ratio 
\be
\chi_C(\omega+i\eta)=\frac{\braket{\sigma_f}_\omega}{h(\omega)}=-\frac{i\beta\Gamma}{\omega+i\Gamma}
\ee
so that $\chi''_C(\omega+i\eta)/\beta\omega=-\pi\delta_\Gamma(\omega)$, where
\be
\delta_\Gamma(\omega)\equiv\frac{\Gamma/\pi}{\omega^2+\Gamma^2}.\label{eq60}
\ee
In time domain, the retarded function is $\chi^R_C(t)=\theta(t)\chi_0\Gamma e^{-\Gamma t}$.
\subsubsection{Markovian-Gaussian formula, Eq.\,\pref{eqMG}}
The retarded charge susceptibility can be obtained from the imaginary-time function $\chi_C(\tau)=-e^{i\Gamma\tau}$. Inserting this into Eq.\,\pref{eqGaussian} and doing the integral we find
\bea
G_d(\tau)&=&-e^{\alpha-\omega_0\tau+i\alpha\Gamma \tau}e^{-\alpha\exp[i\Gamma\tau]}
\eea
where $\alpha=x^2$. We can expand the last exponent as 
$e^{-\alpha\exp[i\Gamma\tau]}=\sum_ne^{in\Gamma\tau}f_n(\alpha)$ where the coefficients $f_n(\alpha)\equiv\sum_mJ_{m}(-\alpha)I_{n-m}(-\alpha)$ are given in terms of ordinary $J_m(x)$ and modified $I_m(x)$ Bessel functions. $f_n(\alpha)$ are real and they are zero for $n<0$. Fourier transforming and analytical continuation gives Eq.\,\pref{eqMG}.

\subsection{Matsubara sum}
We start from $\Sigma_d(\tau)=-\lambda^2G_d(\tau)\chi_C(\tau)$ and do a Fourier transform to write
\be
\Sigma_d(i\omega_n)=-\frac{\lambda^2}{\beta}\sum_mG_d(i\omega_m)\chi_C(i\omega_n-i\omega_m).\label{eqSelfd}
\ee
where $\omega_n=(2n+1)\pi/\beta$ are fermionic Matsubara frequencies. Following the standard procedure, the summation on fermionic Matsubara frequencies is written as a contour integral around the poles of $f(z)$ and the contour is deformed to move the integration parallel to the real frequency axis. The poles of $G_d(z)$ and $\chi_C(i\omega_n-z)$ are along $\im{z}=0$ and $\im{z}=\omega_n$, respectively. Therefore, after analytical continuation of $\Sigma_d(i\omega_n)$ to real frequency and taking the imaginary part, we obtain
\bea
\Sigma''_d(\omega)={\lambda^2}\intinf{\frac{d\omega'}{\pi}}&\Big[&f(\omega')G_d''(\omega')\chi_C''(\omega-\omega')\nonumber\\
&&\hspace{0cm}-n_B(-\omega')G''_d(\omega-\omega')\chi_C''(\omega')\Big].\qquad\label{eq66}
\eea
The part proportional to $f(\omega)$ can be safely dropped in the $T\ll\omega_0$ limit. Assuming that $G_d(\omega+i\eta)$ is the bare propagator $g_d(\omega+i\eta)=[\omega-\omega_0+i\eta]^{-1}$ we arrive at Eq.\,\pref{eqOurSelf}.
\subsubsection{Self-consistent calculation}
Eq.\,\pref{eq66} can be used iteratively for a self-consistent calculation together with the Dyson equaiton $G_d(z)=[z-\omega_0-\Sigma_d(z)]^{-1}$. To get a rough idea of the effect of this self-consistent calculation, we give the $d$ level a natural linewidth and a wavefunction renormalization factor by approximating $G_d(\omega+i\eta)\approx Z[\omega-\omega_0+i\Lambda]^{-1}$. Inserting this into Eq.\,\pref{eq66} leads to
\be
\Sigma''_d(\omega_0+\omega)=-\lambda^2Z\intinf{d\omega'}\delta_\Lambda(\omega-\omega')n_B(-\omega')\chi_C''(\omega').\nonumber
\ee
Therefore, we see that applying the self-consistency leads to broadening of the features in the self-energy. As we showed in the main part of the paper, the self-consistency requires $\Lambda/\Gamma\sim x^2$ and therefore, the slow fluctuation case $x\gg 1$, is affected more by the self-consistent calculations than the fast fluctuation case.
\subsection{Vertex correction}
\begin{figure}[h!]
\includegraphics[width=1\linewidth]{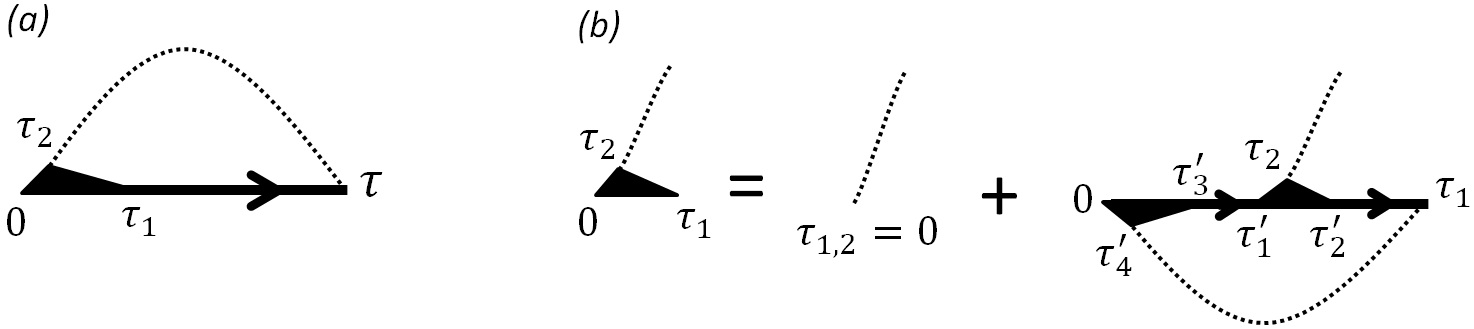}
\caption{\small (a) Diagrammatic representation of Eq.\,\pref{eq57}. (b) Diagrammatic representation of Eq.\,\pref{eq58}, assuming that there is no interaction vertices between charge fluctuations in $H_{\rm rest}$.\label{fig5}}
\end{figure}

Taking into account the vertex correction, Eq.\,\pref{eqSelfd} becomes
\be
\Sigma_d(i\omega_n)=-\frac{\lambda^2}{\beta}\sum_{i\nu_p}\chi_C(i\nu_m)G_d(i\omega_n-\nu_p)\Pi(i\omega_n,i\nu_p),\label{eq57}
\ee
where $\nu_p=2\pi p/\beta$ are bosonic Matsubara frequencies. 
In time-domain this is
\bea
\Sigma_d(\tau)&=&-\lambda^2\int_0^\tau d\tau_1\int_0^{\tau_1}d\tau_2
\chi_C(\tau-\tau_2)\times\nonumber\\
&&\hspace{3.5cm}
G_d(\tau-\tau_1)\Pi(\tau_1,\tau_1-\tau_2),\nonumber
\eea
and assuming that there are no interaction vertices for the charge fluctuation in $H_{\rm rest}$, the vertex function $\Pi$ obeys the equation
\bea
\Pi(\tau_1,\tau_1-\tau_2)&=&\delta(\tau_1)\delta(\tau_2)
\nonumber\\
&&\hspace{-2cm}{-}\lambda^2\int_0^{\tau_2}{d\tau'_1}\int_{\tau_2}^{\tau_1}{d\tau'_2}\int_0^{\tau'_1}{d\tau'_3}\int_0^{\tau'_3}{d\tau'_4}\nonumber\\&&\hspace{-1.5cm}
\Big[G_d(\tau'_1-\tau'_3)G_d(\tau_1-\tau'_2)\chi_C(\tau_1-\tau'_4)\times
\nonumber\\&&\hspace{0cm} 
\Pi(\tau'_3,\tau'_3-\tau'_4)\Pi(\tau'_2-\tau'_1,\tau'_2-\tau_2)\Big]\qquad\label{eq58}
\eea
This is shown diagrammatically in Fig.\,\pref{fig5}.

The leading correction is obtained from the first iteration [shown in Fig.\,(2d)]
\bea
&&\Sigma_d(\tau)=-\lambda^2G_d(\tau)\chi_C(\tau)\nonumber\\
&&+\lambda^4\int_0^\tau d\tau_1\int_0^{\tau_1}d\tau_2\Big[\chi_C(\tau_1)\chi_C(\tau-\tau_2) G_d(\tau-\tau_1)\times \nonumber\\
&&\hspace{5.6cm}G_d(\tau_2)G_d(\tau_1-\tau_2)\Big].\nonumber
\eea
Assuming that $G_d(\tau)=g_d(\tau)$ is the bare propagator and we use the forward-time approximation, the second term becomes
\bea
&&\lambda^4g_d(\tau)\int_0^\tau{d\tau_1}\int_0^{\tau_1}d\tau_2\chi_C(\tau_1)\chi_C(\tau-\tau_2)\nonumber
\\&&\hspace{3cm}
\approx
\lambda^4g_d(\tau)\chi_C(\tau)\int_0^\tau{d\tau_1}\int_0^{\tau_1}d\tau_2\chi_C(\tau_1),\nonumber
\eea
where we have approximated $\chi_C(\tau-\tau_2)\approx\chi_C(\tau)+\cdots$. Assuming that $\chi_C(\tau)$ has a time-scale of $1/\Gamma$, we obtain
\be
\Sigma_d(\tau)=-{\lambda^2}(1-\lambda^2/\Gamma^2)g_d(\tau)\chi_C(\tau),
\ee
resulting in the leading vertex correction $\lambda^2\to\lambda^2(1-x^2)$. Again we see that the fast fluctuation case ($x\ll 1$) is not affected by the vertex corrections. In the opposite limit of fast fluctuations, the result is negative suggesting the break down of the leading order calculation.


\begin{thebibliography}{50}
\bibitem{footnote1}
The linear conductance through a quantum dot is $
G=\alpha\frac{e^2}{h}\int{d\omega[f'(\omega)]G''_d(\omega+i\eta)}$, 
where $\alpha$ is a measure of coupling symmetry of the dot to the source and drain. At low temperature, $G\propto G''(\omega+i\eta)$ and if the dot is in the weak-coupling (to the leads) regime, the system is again described by the Hamiltonian of Eq.\,\pref{eq1} but with $\omega_0\approx 0$. 
Therefore, in this case the forward-time propagator is not a good approximation, and the full Eq.\,\pref{eqOurSelf} has to be used.
\end{thebibliography}
\end{document}